\documentclass{ws-ijmpcs}

\usepackage{color,amsbsy,amssymb,latexsym,amsfonts, amsmath,cancel}
\usepackage{braket}
\usepackage{mathrsfs}
\usepackage{graphicx}
\numberwithin{equation}{section}
\newcommand{\bel}[1]{\begin{equation}\label{#1}}                     
\newcommand{\bal}[1]{\begin{eqnarray}\label{#1}}   
\newcommand{\be}{\begin{equation}}               
\newcommand{\ba}{\begin{eqnarray}}           
\newcommand{\ee}{\end{equation}}
\newcommand{\ea}{\end{eqnarray}}

\newcommand{\bea}{\begin{equation}}
\newcommand{\eea}{\end{equation}}

\begin{document}

%\begin{flushright}
%OCU-PHYS
%\end{flushright}

\markboth{H. Itoyama and N. Maru}
{D-term Dynamical SUSY Breaking}

%%%%%%%%%%%%%%%%%%%%% Publisher's Area please ignore %%%%%%%%%%%%%%%
%
\catchline{}{}{}{}{}
%
%%%%%%%%%%%%%%%%%%%%%%%%%%%%%%%%%%%%%%%%%%%%%%%%%%%%%%%%%%%%%%%%%%%%

\title{D-term Dynamical SUSY Breaking
\footnote{This talk was given by N.M.}
}

\author{Hiroshi Itoyama
%\footnote{
%Typeset names in 8 pt roman, uppercase. Use the footnote to indicate the
%present or permanent address of the author.}
}

\address{
Department of Mathematics and Physics, Graduate School of Science\\
Osaka City University  and \\
Osaka City University Advanced Mathematical Institute (OCAMI) \\
3-3-138, Sugimoto, Sumiyoshi-ku, Osaka, 558-8585, Japan \\
%\footnote{State completely without abbreviations, the
%affiliation and mailing address, including country. Typeset in 8 pt
%italic.}\\
itoyama@sci.osaka-cu.ac.jp}

\author{Nobuhito Maru}

\address{Department of Physics, and Research and Education Center for Natural Sciences, \\
Keio University, Yokohama, 223-8521, Japan\\
maru@phys-h.keio.ac.jp}

\maketitle

%\begin{history}
%\received{Day Month Year}
%\revised{Day Month Year}
%\end{history}

\begin{abstract}
We consider an ${\cal N} =1$ supersymmetric $U(N)$ gauge theory with an adjoint chiral multiplet. 
By developing a self-consistent Hartree-Fock approximation to the leading order 
  which is reminiscent of that of the BCS/NJL in the superconductivity/chiral symmetry, 
  we show that the ${\cal N} =1$ supersymmetry is spontaneously broken, 
  giving a mixed Majorana-Dirac mass term for gaugino due to the nonvanishing $D$-term VEV 
  and $F$-term one induced by $D$-term. 
\end{abstract}

\ccode{PACS numbers: 12.60.Jv, 11.30.Qc}

\section{Introduction}
Supersymmetry (SUSY) is one of the attractive scenarios as the solution to the hierarchy problem, 
but it must be spontaneously broken since sparticles have not been observed so far. 
Two kinds of order parameters of SUSY breaking are known. 
One is a nonvanishing vacuum expectation value (VEV) of F-term in the chiral multiplet, 
and the other is that of D-term in the vector multiplet.  
In the case of F-term SUSY breaking, 
there are the mechanisms generating the nonvanishing VEV of F-term 
both at tree level\cite{Oraif} and at quantum levels.  
To solve the hierarchy problem, SUSY breaking scale should be around TeV scale, 
which requires SUSY to be dynamically broken. 
In SUSY case, this implies that SUSY is broken by nonperturbative effects due to nonrenormalization theorem\cite{GRS}. 
Instantons, and gaugino condensation etc. play an importnat role in generating nonperturbative superpotential 
which causes SUSY breaking. 
Such a mechanism has been well studied so far\cite{Witten}\cdash\cite{Veneetal}. 
On the other hand, %apart from SUSY breaking by a Fayet-Iliopoulos D-term,
D-term SUSY breaking at tree level\cite{FI} is well known, 
but D-term SUSY breaking beyond the lowest order of perturbation theory has not been studied 
so extensively\footnote{Explicit models of displaying dynamical SUSY breaking 
where both F-terms and D-terms are generated has been discussed.\cite{DandF} 
The idea of using a Nambu-Jona-Lasinio type of approach to assess dynamical 
SUSY breaking has already been proposed and studied in the literature.\cite{SUSYNJL}
} 
in spite that it is in principle possible since the nonrenormalization theorem does not apply in this case. 

We propose a new mechanism of dynamical SUSY breaking 
where the nonvanishing VEV of D-term is generated in the sense of self-consistent Hartree-Fock approximation\cite{IM} 
as in the theory of superconductivity\cite{BCS} and chiral symmetry.\cite{NJL} 
We have explicitly derived a gap equation for D-term in a $U(N)$ gauge theory with an chiral adjoint multiplet 
and have indeed shown the existence of its nontrivial solution with $\langle D \rangle \ne 0$. 
Although our SUSY breaking vacuum is a local minimum from the fact that a gap equation still has a trivial solution with $\langle D \rangle=0$, 
we show that the decay rate of our vacuum into the true one can be sufficiently small by adjusting parameters of the theory. 
In our mechanism, scalar gluons are assumed to exist in nature, 
which is distinct from the previous proposals on dynamical SUSY breaking 
and expected to provide specific signatures in collider physics.

%%%%%%%%%%
\section{Basic idea}
%%%%%%%%%%

Let us start from a general lagrangian of SUSY $U(N)$ gauge theory with an adjoint chiral multiplet, 
  % (3.1) IMM
     \ba
    {\cal L}
     &=&     
           \int d^4 \theta K(\Phi^a, \bar{\Phi}^a)
%           \frac{\partial {\cal F}(\Phi)}{\partial \Phi} 
+ (gauging)  %\nonumber \\ &&  
        + \int d^2 \theta
           {\rm Im} \frac{1}{2} 
           \tau_{ab}(\Phi^a)
           {\cal W}^{\alpha a} {\cal W}^b_{\alpha}
            \nonumber \\&& 
            + \left(\int d^2 \theta W(\Phi^a)
         + c.c. \right),      
           \label{KtauW}
    \ea
 where $K$ is a K\"ahler potential with its gauging by the gauge group understood, 
 ${\cal W}^a_\alpha$ is field strength superfields, 
 $\tau_{ab}(\Phi^a)$ is a gauge kinetic function of the chiral superfield $\Phi^a$ 
 in the adjoint representation, and $W(\Phi^a)$ is a superpotential. 
 If we take the K\"ahler potntential, the superpotential and the gauge kinetic function of the form
 \ba
 K(\Phi, \bar{\Phi}) &=& {\rm Im} {\rm Tr}~\bar{\Phi} e^{ad V} \frac{\partial{\cal F}(\Phi^a)}{\partial \Phi^a}, \\
 W(\Phi) &=& {\rm Tr}\left( 2e \Phi + m \frac{\partial {\cal F}(\Phi)}{\partial \Phi} \right), \\
 \tau_{ab}(\Phi) &=& \frac{\partial {\cal F}(\Phi)}{\partial \Phi^a \partial \Phi^b},
 \ea
this model is promoted to an ${\cal N}=2$ supersymmetric theory 
and is known to exhibit the partial spontaneous breaking of ${\cal N}=2$ supersymmetry to ${\cal N}=1$ 
at the tree level\cite{APT}\cdash\cite{FIS}.   
Our mechanism discussed below can be also applied in this case. 

 The fermion bilinears made of the gaugino $\lambda^a$ and the matter fermion $\psi^a$ 
 %which are referred to as ${\cal N}=2$ gauginos in this letter, 
 are obtained from the second and the third line of Eq.~(\ref{KtauW}): 
%FORMULA7 draft
\ba
-\frac{1}{2} (\lambda^a~\psi^a) 
\left(
\begin{array}{cc}
0 & -\frac{\sqrt{2}}{4} \tau_{abc} D^b \\
-\frac{\sqrt{2}}{4} \tau_{abc} D^b & \partial_a \partial_c W \\
\end{array}
\right) 
\left(
\begin{array}{c}
\lambda^c \\
\psi^c \\
\end{array}
\right) + (c.c.), 
\label{Majorana-Dirac}
\ea 
 where $\tau_{abc} \equiv \partial_c \tau_{ab}$ implies the derivative of $\tau_{ab}$ with respect to $\Phi^c$.
 Note that the nonvanishing value of $\tau_{abc}$ ensures the coupling of the auxiliary field $D^a$
   to the fermionic bilinears while there is no bosonic  counterpart in the lagrangian.
 Let us assume that $\tau_{ab}$ is obtained as the second derivatives of a trace function $f(\Phi^a)$.
  The nonvanishing VEVs are $\langle \tau_{0aa} \rangle$.
 The holomorphic and nonvanishing part of the mass matrix is
\ba
  M_{a, a} \equiv
\left(
\begin{array}{cc}
0 & -\frac{\sqrt{2}}{4} \langle \tau_{0aa} D^0 \rangle \\
-\frac{\sqrt{2}}{4} \langle \tau_{0aa} D^0 \rangle & \langle \partial_a \partial_a W \rangle \\
\end{array}
\right) 
\label{Fmassmat}
\ea 
 to each generator. 
 Upon diagonalization, this has two unequal and nonvanishing eigenvalues
  provided  $ \langle D^0 \rangle \neq 0$ and
 $\langle \partial_a \partial_a W \rangle \neq 0$.
\ba
{\bf \Lambda}_a^{\pm} = \frac{1}{2} m_a 
\left[
1 \pm \sqrt{1+ \Delta^2} \right], \quad m_a \equiv \langle g^{aa} \partial_a \partial_a W \rangle, \quad 
\Delta^2 \equiv \frac{\langle \tau_{0aa}D^0 \rangle^2}{2 \langle \partial_a \partial_a W \rangle^2}. 
\label{eigenvalue}
\ea 
  In this case, the gaugino and matter fermion receive masses of mixed Majorana-Dirac type and are split.
 This observation generalizes the proposal of Ref.~\refcite{FNW} where the masses are
  of a pure Dirac type and the gaugino and matter fermion are degenerate while the supersymmetry is broken.
 
The next issue to be investigated is to determine the value of $\langle D^0 \rangle$.  
Note that the equation of motion for $D^0$
  tells us that the condensation of the Dirac bilinears is responsible 
  for the nonvanishing order parameter:
$ \langle D^{0} \rangle
    =     - \frac{1}{2 \sqrt{2}} \langle g^{00} 
           \left( \tau_{0cd}\psi^d \lambda^c 
         + \bar{\tau}_{0cd} \bar{\psi}^d \bar{\lambda}^c \right) \rangle$.
Inspired by this property similar to NJL model, 
we expect that the value of $\langle D^0 \rangle$ is given by the nontrivial solution to the gap equation for D-term.

%%%%%%%%%%%%%%%%%%%%%%%%%%%%%%%%%
\section{Self-consistent Hartree-Fock Approximation and DDSB}
%%%%%%%%%%%%%%%%%%%%%%%%%%%%%%%%%

Our goal here is to describe a procedure to construct
the effective potential in a self-consistent Hartree-Fock approximation which
permits  the auxiliary fields $D^0$ to receive a nonvanishing VEV, and thereby give 
gaugino and matter fermion a mixed mass term of Majorana-Dirac type.  
We will carry this out to one-loop order. 
The procedure gives an optimal value for the scalar VEV as well to this order.
 
%%%%%%%%%%%%%%%%%%%%%%%%%%%%%%
\subsection{One-loop contribution to the effective potential}
%%%%%%%%%%%%%%%%%%%%%%%%%%%%%%
 Let us consider the effective potential to one-loop order:
\ba 
V = V_{{\rm tree}} + V_{{\rm 1-loop}} + V_{\rm c.t.}.  %V_{tree} = V^{(D)} + V^{(sup)}
\ea
$V_{\rm c.t.}$ is a counter term which will be introduced in the next subsection.
The potential at the tree level is given by
% FORMULA 1 draft
\ba
&&V_{{\rm tree}} = V^{(D)} + V^{({\rm sup})}, 
\ea
where $V^{(D)} = -  \frac{1}{2}g_{ab} D^a D^b, V^{({\rm sup})} = g^{ab} \partial_a W \overline{\partial_b W}$.
%\ea
 Note that the minus sign in front of $V^{(D)}$ is not erraneous.
 The one-loop effective potential $V_{{\rm 1-loop}}$ will turn out to be essentially a variant of
  the Coleman-Weinberg potential.
 
It is now clear that the entire contribution to the 1PI vertex function $i \Gamma_{\rm{1-loop}}$
 is
\ba
 \int d^4x \sum_a | m_{a} |^4 \int \frac{d^4 \ell^{\mu}}{(2\pi)^4} \ln
 \left[ \frac{(\lambda^{(+)2} - \ell^2 - i \epsilon) (\lambda^{(-)2} - \ell^2 - i \epsilon)}
 {(1 - \ell^2 - i \epsilon) ( - \ell^2 - i \epsilon)}
  \right].
 \ea 
There are methods available to regulate and evaluate this expression.  For instance, one can make
exploit the integral
\ba
 \log \frac{a}{b} = - \int_0^{\infty} \frac{dt}{t} (e^{-at} - e^{-bt})
\ea
 to obtain the one-loop contribution to the effective potential:
\ba
V_{{\rm 1-loop}} = \frac{1}{16 \pi^2} \sum_a |m_a|^4 \int_{0}^{\infty}
 \frac{dt}{t^3} \left( e^{- \lambda^{(+)2} t} + e^{- \lambda^{(-)2} t} - e^{-t} -1 \right).
\ea
 Analytically continuing $3$ to $1+d/2$ in the case of $d$ dimensional integral,  we obtain 
 \ba
 V_{{\rm 1-loop}} = \sum_a |m_a|^4
    \frac{1}{32\pi^2} \left[ A(d) \left( \Delta^2 + \frac{1}{8} \Delta^4 \right) - 
  \lambda^{(+)4} \log \lambda^{(+)2}   - \lambda^{(-)4}  \log \lambda^{(-)2} \right], \nonumber \\
  \ea
  where
 \ba
  A(d) = \frac{3}{4} - \gamma +\frac{1}{2-d/2}.
 \ea
  
%%%%%%%%%%%%%%%%%%%%%%%%%%%%%%%%%%%%%%%%%%
\subsection{Subtraction of UV divergence and the effective potential up to one-loop}
%%%%%%%%%%%%%%%%%%%%%%%%%%%%%%%%%%%%%%%%%%
 The theory under consideration is nonrenormalizable at least by power counting
 but we would still like to isolate UV infinities and to subtract them by adding a supersymmetric counter term to one-loop effective potential 
 \ba
V_{{\rm c.t.}} = -{\rm Im} \frac{\Lambda}{2} \int d^2 \theta {\cal W}^{\alpha a}{\cal W}^a_\alpha 
= -{\rm Im} \frac{\Lambda}{2} (D^0)^2. 
\ea
Now the part of the one-loop effective potential which contains $\Delta$ reads
 \ba
  V_{{\rm 1-loop}}^{(D)} &=& V^{(D)} + V_{{\rm c.t.}} + V_{{\rm 1-loop}}  \nonumber \\ 
   &=& \sum_a | m_{a} |^4 \left[ - \beta \Delta^2 - \Lambda_{{\rm res}} \Delta^2 \right. \nonumber \\
   &+& \left. \frac{1}{32 \pi^2} \left\{ A(d) \left( \Delta^2 + \frac{1}{8} \Delta^4 \right)
     -   \lambda^{(+)4} \log \lambda^{(+)2}   - \lambda^{(-)4}  \log \lambda^{(-)2} \right\}
  \right],
 \ea
 where
 \ba
  \beta = \frac{ \langle g_{00} \rangle | \langle \partial_a \partial_a W \rangle |^2}{\sum_a | m_{a} |^4 |\langle \tau_{0aa} \rangle|^2}, \;\;
  \Lambda_{{\rm res}} = \frac{ ({\rm Im}) \Lambda |\langle \partial_a \partial_a W \rangle|^2}{\sum_a | m_{a} |^4 |\langle \tau_{0aa} \rangle|^2}.
 \ea
 We will absorb the infinity in $A(d)$ into $\Lambda_{{\rm res}}$ by imposing one condition,
   which is, for instance,
 \ba
  \frac{1}{\sum | m_{a} |^4} \left. \frac{\partial^2 V}{(\partial \Delta)^2} 
\right|_{\Delta=0} = 2c,
 \ea
  where $c$ is a fixed non-universal number.
  We finally obtain
 \ba
V^{(D)}_{{\rm 1-loop}}  
   &=&  \sum_a |m_a|^4 
   \left[
   \left( c + \frac{1}{64 \pi^2} \right)  \Delta^2  + \Lambda_{{\rm res}}^{\prime} 
   \frac{\Delta^4}{8}
  \right.              \nonumber \\ 
   && \left. 
   - \frac{1}{32 \pi^2} 
  \left(  \lambda^{(+)4} \log \lambda^{(+)2}   + \lambda^{(-)4}  \log \lambda^{(-)2} \right) \right], 
 \label{1looppot}
 \ea
 where $\Lambda_{{\rm res}}^{\prime} \equiv c + \beta + \Lambda_{{\rm res}} +\frac{1}{64\pi^2} = \frac{1}{32\pi^2} A(d)$.

%%%%%%%%%%%%%%%%%%%%%%%%%%%%%%%%%%%
\subsection{Gap equation and vacuum condition at one-loop level}
%%%%%%%%%%%%%%%%%%%%%%%%%%%%%%%%%%%
The gap equation can be obtained from the stationary condition of the potential at one-loop (\ref{1looppot}) with respect to $\Delta$. 
\ba
0 &=& \frac{\partial V^{(D)}_{{\rm 1-loop}}}{\partial \Delta} \nonumber \\
&=& \Delta \left[ 2 \left( c+\frac{1}{64\pi^2} \right) + \frac{\Lambda'_{{\rm res}}}{2} \Delta^2 
%\right. \nonumber \\
%&& \left.  
- \frac{1}{32\pi^2} \frac{1}{\sqrt{1+\Delta^2}} 
\left\{ (\lambda^{(+)})^3 (2 \log(\lambda^{(+)})^2 + 1) \right. \right. \nonumber \\
&& \left. \left. - (\lambda^{(-)})^3 (2 \log(\lambda^{(-)})^2 + 1) \right\} 
\right]. 
\ea
The gap equation still has a trivial solution $\Delta=0$. 
Now our interest is whether the nontrivial solution $\Delta \ne 0$ exists or not. 
Let us solve the equation approximately since it is transcendental. 
First consider the case where the D-term VEV is very small, $\Delta^2 \ll 1$. 
Noting that $\lambda^{(+)} \simeq 1+\Delta^2/4, \lambda^{(-)} \simeq 1-\Delta^2/2$ in this case, 
the gap equation can be approximated as 
\ba
2 \left( c+\frac{1}{64\pi^2} \right) + \frac{\Lambda'_{{\rm res}}}{2} \Delta^2 \simeq \frac{1}{32\pi^2} \left( 1 + \frac{5}{4}\Delta^2 \right). 
\label{apgap1}
\ea
If $c > 0$, then we have no solution of (\ref{apgap1}) because of $\Lambda'_{{\rm res}}$. 
If $c < 0$, then we have a solution $\Delta^2 \simeq -4c/(\Lambda'_{{\rm res}}-\frac{5}{64\pi^2})$. 

Next consider the case where the D-term VEV is very large, $\Delta^2 \gg 1$. 
Noting that $\lambda^{(\pm)} \simeq \pm \Delta/2$ in this case, 
the gap equation can be approximated as 
\ba
c+\frac{1}{64\pi^2} + \Lambda'_{{\rm res}} \left( \frac{\Delta}{2} \right)^2 
\simeq \frac{1}{32\pi^2} \left( \frac{\Delta}{2} \right)^2 \log \left( \frac{\Delta}{2} \right)^2, 
\label{apgap2}
\ea
which has a unique nontrivial solution $\Delta \ne 0$ if $\Lambda'_{{\rm res}} > 0$. 

The fact that the nontrivial solution to the gap equation exists can be also checked numerically.  
%%%%%%%%%%%%%%%%%%%%%%%
\begin{figure}[htbp]
 \begin{center}
  \includegraphics[width=70mm]{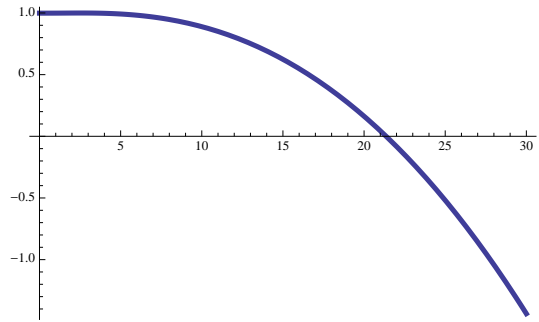}
 \end{center}
 \caption{The plot of the quantity $\partial V^{(D)}_{{\rm one-loop}}/(\Delta \partial \Delta)$ as a function of $\Delta$. 
 A particular set of parameters $c+\frac{1}{64\pi^2}=1, \Lambda_{{\rm res}}'/8=0.001$ is chosen as an illustration.}
 \label{fig:one}
\end{figure}
%%%%%%%%%%%%%%%%%%%%%%%
In Figure \ref{fig:one}, the quantity $\partial V^{(D)}_{{\rm one-loop}}/(\Delta \partial \Delta)$ is plotted as a function of $\Delta$. 
Obviously, a unique zero point can be seen, which implies the existence of a nontrivial solution to the gap equation.  
Although a particular set of parameters is chosen in the Figure \ref{fig:one} as an illustration, 
we have checked that a nontrivial solution indeed exists in a wide range of parameters.  
Therefore, we conclude that SUSY is broken by dynamically generated VEV of D-term in a self-consistent Hartree-Fock approximation.

%%%%%%%%%%%%%%%%%%%%%%%%%%%%%%%%%%%%%%%%%
\section{Nonvanishing $\langle F^0 \rangle$ induced by $\langle D^0 \rangle \ne 0$}
%%%%%%%%%%%%%%%%%%%%%%%%%%%%%%%%%%%%%%%%%
In the above analysis, we have shown that SUSY is dynamically broken by the nonvanishing VEV of D-term 
to the leading order in the Hartree-Fock approximation. 
However, this is not sufficient to discuss phenomenological applications to the observable sector appropriately. 
We have to note the fact that nonvanishing VEV of D-term induces that of F-term in general. 

Let us see how this happens in our case. 
The VEV of the scalar fields get shifted in the vacuum of nonvanishing $\Delta$ 
and then the F-term also develops a nonvanishing VEV as a result of the vacuum condition. 
\ba
\langle \delta V \rangle = 0 \Rightarrow
|\langle F^0 \rangle|^2 + \frac{m_0}{\langle g^{00} \partial_0 g_{00} \rangle} \langle F^0 \rangle 
+ \frac{1}{2} \langle D^0 \rangle^2 
+ 2 \langle g^{00} \rangle \langle V_{{\rm 1-loop}} \rangle =0. 
\label{vcond1}
\ea
Combining with the condotion $\langle \bar{\delta} V \rangle=0$, we further obtain
\ba
\frac{m_0}{\langle g^{00} \partial_0 g_{00} \rangle} \langle F^0 \rangle 
= \frac{m_0^*}{\langle \overline{g^{00} \partial_0 g_{00}} \rangle} \langle \bar{F}^0 \rangle. 
\label{vcond2}
\ea
These two relations (\ref{vcond1}) and (\ref{vcond2}) determine the nonvanising F-term induced by the nonvanishing D-term.

%%%%%%%%%%%%%%%%%%%%%%%%
\section{The metastability of our vacuum}
%%%%%%%%%%%%%%%%%%%%%%%%
Combining the two facts that the trivial solution $\Delta=0$ of the gap equation is also a solution 
and the energy in rigid SUSY theories is positive semi-definite leads us that our SUSY breaking vacuum is a local minimum. 
For our mechanism to be viable, we have to show that our SUSY breaking vacuum is sufficiently long-lived 
during the decay into the true vacuum with $\Delta=0$; in other words, 
the lifetime of our vacuum must be much longer than the age of universe. 
Taking into account the nonvanishing F-term VEV as well discussed in the previous section, 
we carry out an order estimate of the lifetime of our SUSY breaking vacuum. 
Neglecting ${\cal O}(1)$ quantities, we have
\ba
\frac{1}{2} \langle D^0 \rangle^2 \sim {\cal O}(m_0^2 \Lambda^2), 
\quad \langle V_{{\rm 1-loop}} \rangle \sim {\cal O} \left(\frac{\alpha}{4\pi} N^2 m_0^4 \right)  
\sim {\cal O} \left(m_0^4 \right) 
\label{VEV1}
\ea
where $\Lambda$ is a cutoff scale. 
Plugging these VEVs into eq.~(\ref{vcond1}) leads to 
\ba
\langle F^0 \rangle \sim {\cal O}(m_0 \Lambda)
\ea
provided $m_0 \ll \Lambda$. 

The decay rate of our vacuum to the true one is controlled by the factor $\exp[ - |\langle \Delta \phi \rangle|^4/\langle \Delta V \rangle]$ as seen in Ref.\refcite{ISS},  
where $\langle \Delta \phi \rangle, \langle \Delta V \rangle$ are the shift of the scalar field VEV, the potential height between two vacua. 
These two quantities are estimated as follows. 
\ba
&&\langle F^0 \rangle %= \langle F^0_{\bar{0}} \rangle \langle \Delta \overline{\phi}^0 \rangle 
= - \langle g^{00}\rangle \langle \overline{\partial_0 \partial_0 W} \rangle  \langle \Delta \overline{\phi}^0 \rangle 
=- m_0 \langle \Delta \overline{\phi}^0 \rangle \Rightarrow \langle \Delta \overline{\phi}^0 \rangle \sim {\cal O}(\Lambda), \\
&&\langle \Delta V \rangle = |\langle F^0 \rangle|^2 + \frac{1}{2} \langle D^0 \rangle^2 + \langle V_{{\rm 1-loop}} \rangle \sim {\cal O}(m_0^2 \Lambda^2). 
\ea
Using these results, the requirement of the longevity of our metastable vacuum is given by the condition  
\ba
\frac{| \langle \Delta \phi^0 \rangle|^4}{\langle \Delta V \rangle} \sim {\cal O}\left( \frac{\Lambda^2}{m_0^2} \right) \gg 1,
\ea
which is always satisfied as long as $m_0 \ll \Lambda$. 

%%%%%%%%%%%
\section{Application}
%%%%%%%%%%%
In the phenomenological model building, 
our mechanism of SUSY breaking can be applied to the model of SUSY breaking mediation\cite{FNW}. 
Consider an ${\cal N}=2$ extension of the gauge sector and ${\cal N}=1$ matter sector of the minimal SUSY Standard Model (MSSM). 
The reasons why the MSSM matter sector is not extended to ${\cal N}=2$ are that we need chiral fermions for quarks and leptons 
and we would like to avoid Landau pole of QCD gauge coupling. 
The ${\cal N}=2$ extention of the gauge sector leads to the existence of the scalar gluons, 
which are expected to provide distinct signatures of collider physics and so on as a prediction of our mechanism. 
As for the gauge group, we take $G' \times G_{{\rm SM}}$ 
where $G'(G_{{\rm SM}})$ are the hidden (SM) gauge group, respectively. 
$G'$ includes an overall $U(1)$ gauge group responsible for the dynamical SUSY breaking. 

The key ingredient to work our mechanism of SUSY breaking is the dimension five operator in the gauge kinetic term, 
\ba
\int d^2 \theta \tau_{abc}(\Phi) \Phi^c_{{\rm SM}} {\cal W}^{'a \alpha} {\cal W}^{b}_{\alpha{\rm SM}} 
\Rightarrow \tau_{abc}(\langle \Phi \rangle) \langle D^{'a} \rangle \psi^{c \alpha}_{{\rm SM}} \lambda^{b}_{\alpha{\rm SM}} 
\ea
which is a source of Dirac gaugino mass by the nonvanishing D-term VEV. 
As discussed in section 4, 
the nonvanising F-term is also induced by that of D-term, 
which implies that the $SU(N)$ gaugino mass of the form such as (\ref{eigenvalue}) is no longer valid. 
In fact, the holomorphic part of the fermion mass matrix to the leading order is modified as
\ba
{\cal L}_{{\rm mass}}^{({\rm holo})} &=& -\frac{1}{2} \langle g_{0a,a} \rangle \langle \bar{F}^0 \rangle \psi^a \psi^a 
+ \frac{i}{4} \langle {\tau}_{0aa} \rangle \langle F^0 \rangle \lambda^a \lambda^a
-\frac{1}{2} \langle \partial_a \partial_a W \rangle \psi^a \psi^a 
\nonumber \\
&&+ \frac{\sqrt{2}}{4} %\left( 
\langle {\tau}_{0aa} \rangle \psi^a \lambda^a 
%+ \langle \bar{{\tau}_{0aa} \rangle \bar{\psi}^a \bar{\lambda}^a \right) 
\langle D^0 \rangle \nonumber \\
&\equiv& -\frac{1}{2} \sum_{a=1}^{N^2-1} \Psi^{aT} \tilde{M}_{a,a} \Psi^a, \quad 
\Psi^a = 
\left(
\begin{array}{c}
\lambda^a \\
\psi^a \\
\end{array}
\right).  
\ea
The $SU(N)$ gaugino and matter fermion masses are obtained by diagonalizing this modified mass matrix $\tilde{M}_{a,a}$.  

As for gaugino and matter fermions in the $U(1)$ sector, 
the hidden sector where the Nambu-Goldstone fermion resides, the index loop circulates in the one-loop self energy part as well, 
which is, in additon to the above contributions, regarded as the leading contribution to the mass matrix. 
The massless fermion ensured by the theorem is an admixture of $\lambda^{0}$ and $\psi^{0}$.

Once the gaugino masses are generated, 
the SUSY breaking is transmitted to the sfermion masses by 1-loop RGE effects\cite{FNW}. 
\ba
m_{{\rm sfermion}}^2 \simeq \frac{C_i(R) \alpha_i}{\pi} M^2_{\lambda_i} 
\log \left[\frac{m_a^2}{M^2_{\lambda_i}} \right]~(i=SU(3), SU(2), U(1))
\ea
where $M_{\lambda_i}$ are the gaugino masses of the SM, $C_i$ is the quadratic Casimir of representation $R$ 
and $\alpha_i$ are the fine structure constants. 
These sfermion masses are positive and flavor-blind. 
Therefore our sparticle spectrum is free from supersymmetric flavor and CP problems.

%%%%%%%%%%%
\section{Conclusions}
%%%%%%%%%%%
In this talk, we have proposed a new mechanism of dynamical SUSY breaking 
where a nonvanishing VEV of D-term is generated 
at the leading order in the self-consistent Hartree-Fock approximation 
as in the theory of superconductivity and chiral symmetry. 
In our mechanism, scalar gluons is assumed to exist in nature, 
which is distinct from the previous proposals on dynamical SUSY breaking. 
We have explicitly derived a gap equation for D-term in a $U(N)$ gauge theory with an chiral adjoint multiplet 
and have indeed shown the existence of its nontrivial solution with $\langle D \rangle \ne 0$. 

Although our SUSY breaking vacuum is a local minimum from the fact that a trivial solution with $\langle D \rangle=0$ 
is still a solution of the gap equation, we have shown that the decay rate of our vacuum into the true one can be sufficiently small 
as long as the mass of scalar gluons is quite small comparing to the cutoff scale $m_0 \ll \Lambda$.  

An application to phenomenology is briefly discussed. 
Our mechanism is easy to apply to the supersoft SUSY breaking scenario\cite{FNW} 
where only the gauge sector of the MSSM is extended to ${\cal N}=2$ and 
the Dirac gaugino masses are generated by nonvanishing D-term VEV. 
But our case is more involved since the gauginos receive mixed Majorana-Dirac masses 
from the contributions of nonvanishing F-term as well as D-term. 
The flavor-blind sfermion masses are generated by 1-loop RGE effects similar to Ref.~\refcite{FNW}. 
Our prediction is the existence of the scalar gluons, which are expected to provide distinct signatures of collider physics and so on. 
This issue will be studied in future.

\section*{Acknowledgments}

The authors would like to thank the organizers of the conference for providing us the opportunity 
to give a talk on our recent work at the conference. 
Their research is supported in part by the Grant-in-Aid for Scientific Research  
  from the Ministry of Education, Science and Culture, Japan (23540316 (H.~I.), 21244036 (N.~M.))
  and by Keio Gijuku Academic Development Funds (N.~M.).

\appendix

\section{Calculation of one-loop effective potential by cutoff regularization}

In this appendix, we calculate one-loop effective potential for D-term by using the cutoff regularization 
as was done in the NJL model. 
%We will see that the potential becomes negative in most range of parameters, 
%which is inconsistent with the positive semi-definiteness of the energy in SUSY theories. 
%We instead calculated the potential in the main text by using dimensional reduction, 
%a SUSY counter term was added and a renormalization condition was imposed to remove the infinity. 
Introducing the cutoff scale $\Lambda$ normalized by $m_0$, we obtain the following results
\ba
V_{{\rm 1-loop}}^{(D)} &=& -\frac{m_0^4}{32\pi^2} 
\left[
\Lambda^4 \log 
\left(
\frac{
(\Lambda^2 + \lambda^{(+)2})(\Lambda^2 + \lambda^{(-)2})
}
{
(\Lambda^2 + 1) \Lambda^2
}
\right)
+ (\lambda^{(+)2} + \lambda^{(-)2} -1) \Lambda^2 \nonumber \right. \\
&&\left.  -\lambda^{(+)4} \log \left( 1+\frac{\Lambda^2}{\lambda^{(+)2}} \right)
-\lambda^{(-)4} \log \left( 1+\frac{\Lambda^2}{\lambda^{(-)2}} \right)
+ \log (1+\Lambda^2)
\right] \\
&\simeq& 
\left\{
\begin{array}{l}
-\frac{\Delta^2}{32\pi^2} m_0^4 [\Lambda^2 -\log(1 + \Lambda^2)]~(\Delta \ll 1, \Lambda) \\
-\frac{2}{32\pi^2} m_0^4 \left(\frac{|\Delta|}{2} \right)^4 \log \left(\frac{|\Delta|}{2\Lambda} \right)^2~(\Delta \gg 1, \Lambda)\\
-\frac{\Delta^2 \Lambda^2}{32\pi^2} m_0^4
\left[
1-\frac{\Delta^2}{8\Lambda^2} \log \left( \frac{4\Lambda^2}{\Delta^2} \right) 
+ \frac{1}{\Delta^2 \Lambda^2} \log \Lambda^2
\right]~(1 \ll \Delta \ll \Lambda) \\
-\frac{1}{32\pi^2} m_0^4 
\left[
\Lambda^4 \log \left( 1+ \frac{\Delta^4}{16\Lambda^2} \right) 
+ \frac{1}{2}\Delta^2\Lambda^2 \right. \nonumber \\
\left. \hspace*{24mm} - \left( \frac{\Delta^4}{16} \right)^2 \log \left( 1+ \frac{16\Lambda^2}{\Delta^4} \right)
\right]~(\Lambda \ll \Delta \ll 1). 
\end{array}
\right.
\ea
where the last case might be unphysical since D-term VEV $\Delta$ is much larger than the cutoff scale. 
In the above expression, the approximate forms of potential in the various limit of parameters are derived. 
%We can see from them that the potential becomes negative in most range of parameters. 
%Therefore, we regard the cutoff regularization scheme as an inappropriate procedure in our calculation. 

%\begin{thebibliography}{000} %for 3 digits
%\begin{thebibliography}{00}  %for 2 digits

\end{document}